\documentclass[11pt,titlepage]{article}

\usepackage{setspace} 

\usepackage{amsmath}

\usepackage{graphicx}

\usepackage[round,numbers,sort&compress]{natbib} 
\usepackage{xcolor}

\usepackage{placeins}
\usepackage{ulem}
\title{Anomalous versus slowed-down Brownian diffusion in the ligand-binding equilibrium}

\author{H\'edi Soula\thanks{
           Corresponding author.  hedi.soula@insa-lyon.fr}, Bertrand Car\'{e} \\
	EPI Beagle, INRIA Rh\^one-Alpes\\ F-69603, Villeurbanne, France\\
	and Universit\'e de Lyon, Inserm UMR1060\\ F-69621 Villeurbanne, France 
	\and Guillaume Beslon and Hugues Berry\thanks{
           Corresponding author.  hugues.berry@inria.fr}\\
	EPI Beagle, INRIA Rh\^one-Alpes\\ F-69603, Villeurbanne, France\\
	and LIRIS, Universit\'e de Lyon\\ UMR 5205 CNRS-INSA, F-69621, Villeurbanne, France }

\date{}

\begin{document}

\def\longrightharpoonup{\relbar\joinrel\rightharpoonup}
\def\longleftharpoondown{\leftharpoondown\joinrel\relbar}

\def\longrightleftharpoons{
  \mathop{
    \vcenter{
      \hbox{
	\ooalign{
	  \raise1pt\hbox{$\longrightharpoonup\joinrel$}\crcr
	  \lower1pt\hbox{$\longleftharpoondown\joinrel$}
	}
      }
    }
  }
}

\newcommand{\rates}[2]{\displaystyle
  \mathrel{\longrightleftharpoons^{#1\mathstrut}_{#2}}}

\maketitle

\abstract{Measurements of protein motion in living cells and membranes consistently report transient anomalous diffusion (subdiffusion) which converges back to a Brownian motion with reduced diffusion coefficient at long times, after the anomalous diffusion regime. Therefore, slowed-down Brownian motion could be considered the macroscopic limit of transient anomalous diffusion. On the other hand, membranes are also heterogeneous media in which Brownian motion may be locally slowed-down due to variations in lipid composition. Here, we investigate whether both situations lead to a similar behavior for the reversible ligand-binding reaction in 2d. We compare the (long-time) equilibrium properties obtained with transient anomalous diffusion due to obstacle hindrance or power-law distributed residence times (continuous-time random walks) to those obtained with space-dependent slowed-down Brownian motion. Using theoretical arguments and Monte-Carlo simulations, we show that those three scenarios have distinctive effects on the apparent affinity of the reaction. While continuous-time random walks decrease the apparent affinity of the reaction, locally slowed-down Brownian motion and local hinderance by obstacles both improve it. However, only in the case of slowed-down Brownian motion, the affinity is maximal when the slowdown is restricted to a subregion of the available space. Hence, even at long times (equilibrium), these processes are different and exhibit irreconcilable behaviors when the area fraction of reduced mobility changes.}

\clearpage

\section*{Introduction}

The structural elements of living cells (membranes, cytoplasm, nucleus, mitochondria) exhibit disorder, heterogeneity and obstruction typical of poorly-connected media~\citep{Dix2008}. For instance, cell membranes are heterogeneous collections of contiguous spatial domains with various length and time scales (e.g. fences, lipid rafts, caveolae)~\citep{Jacobson2007}, that spatially modulate the diffusion of proteins~\citep{Schnitzer1993,Kenworthy2004,Kusumi2011}. This defines a spatially heterogeneous diffusion problem, with position-dependent diffusion coefficient~\citep{Kenworthy:2004p378,Goodwin:2005p1132,Pucadyil:2006p1150,Fujita:2007p1107,Day:2009p2559}. On the other hand, the movement of biomolecules such as proteins in the membranes of living cells has consistently been reported to exhibit anomalous diffusion, whereby the mean squared displacement scales sub-linearly with time, $\left\langle \mathbf{r}^{2}(t)\right\rangle\propto t^{\alpha}$ with $\alpha < 1$~\cite{Schwille1999,Smith1999,Fujiwara2002,Weigel2011}.
Such anomalous diffusion phenomenon (also coined subdiffusion) is a hallmark of diffusion obstruction by obstacles~\citep{Bouchaud1990} or random walks with heavy-tailed residence time distributions~\citep{Bouchaud1990,Metzler2000} (for the sake of conciseness, we will not consider here fractional Brownian motion as a model of crowding-induced anomalous diffusion~\cite{barkai-phystoday-2012}).

The influences of such deviations from simple Brownian motion on the biochemical reactions that take place in these media are just starting to be explored. The fundamentally heterogeneous spatial organization of the cell membrane is believed to locally favor the oligomerization of membrane receptors and prolong their local residence times, thus affecting signal transduction in the plasma membrane~\cite{Kusumi2011}. But careful investigations by Monte-Carlo simulations hint that complex or counter-intuitive behaviors can generically be expected~\citep{Grecco2011,Soula2012}.

Investigating the effects of anomalous diffusion on the dynamics of simple elementary reactions of biological interest has recently started to attract the interest of several groups, see e.g.~ref. \cite{Hornung2005,Yuste2010,Fedotov2011,Fedotov2012} to cite only a few or ref.~\cite{AvrahamHavlinBook2005} for a book on elementary irreversible reactions. In the case of the binary reaction A+B$\to products$, for instance, anomalous diffusion alters the overall reaction kinetics~\cite{Saxton2002,Bujan-Nunez2004,Kim2009} and may e.g. favor the search of target DNA sequences by transcription factors in the nucleus~\cite{Golding2006,Guigas2008} or reduce the time needed by an enzyme to reach its substrate~\cite{Sereshki2012}. Anomalous diffusion has also been proposed as a key regulator of the spatiotemporal dynamics of Michaelis-Menten enzyme reactions (E+S$\rightleftharpoons$ C$ \to$ E + P)~\cite{Berry2002,Hellmann2012}. 

The case of reversible reactions, such as the ubiquitous ligand-binding equilibrium 
\begin{equation}\label{eq:LR}
{\rm L}+{\rm R} \rates{ k_{\rm on}} { k_{\rm off}} {\rm C}
\end{equation}
(where L is the ligand, R its free receptor, and C the bound complex) received less attention.  Reversible reactions are expected to converge at long times to equilibrium, thus permitting the study of the influence of anomalous or position-dependent diffusion not only on transient regimes but also long-time equilibrium properties. Indeed, from standard mass-action laws (see e.g.~\cite{Lauffenburger93} for a textbook), the concentration of complex C in reaction eq.~\eqref{eq:LR} evolves according to $dC(t)/dt=k_{\rm on} L(t)R(t)-k_{\rm off}C(t)$, where $X(t)$ is the concentration of species X at time $t$. At equilibrium ($dC/dt=0$), assuming that the total amount of $L$ molecules, $L_T$ is much larger than that of $R$, $R_T$, these mass-action laws yield $C_{eq}=L_TR_T/(K_D+L_T)$ where the equilibrium constant $K_{D}=k_{\rm on}/k_{\rm off}$. This defines so-called dose-response curves -- the equilibrium amounts of $C$ for increasing doses of ligand -- with equilibrium constant $K_{D}$ a measure of the reaction affinity (the smaller $K_D$ the larger the affinity).

\begin{figure}[h!]
 \begin{center}
\includegraphics[scale=1.0]{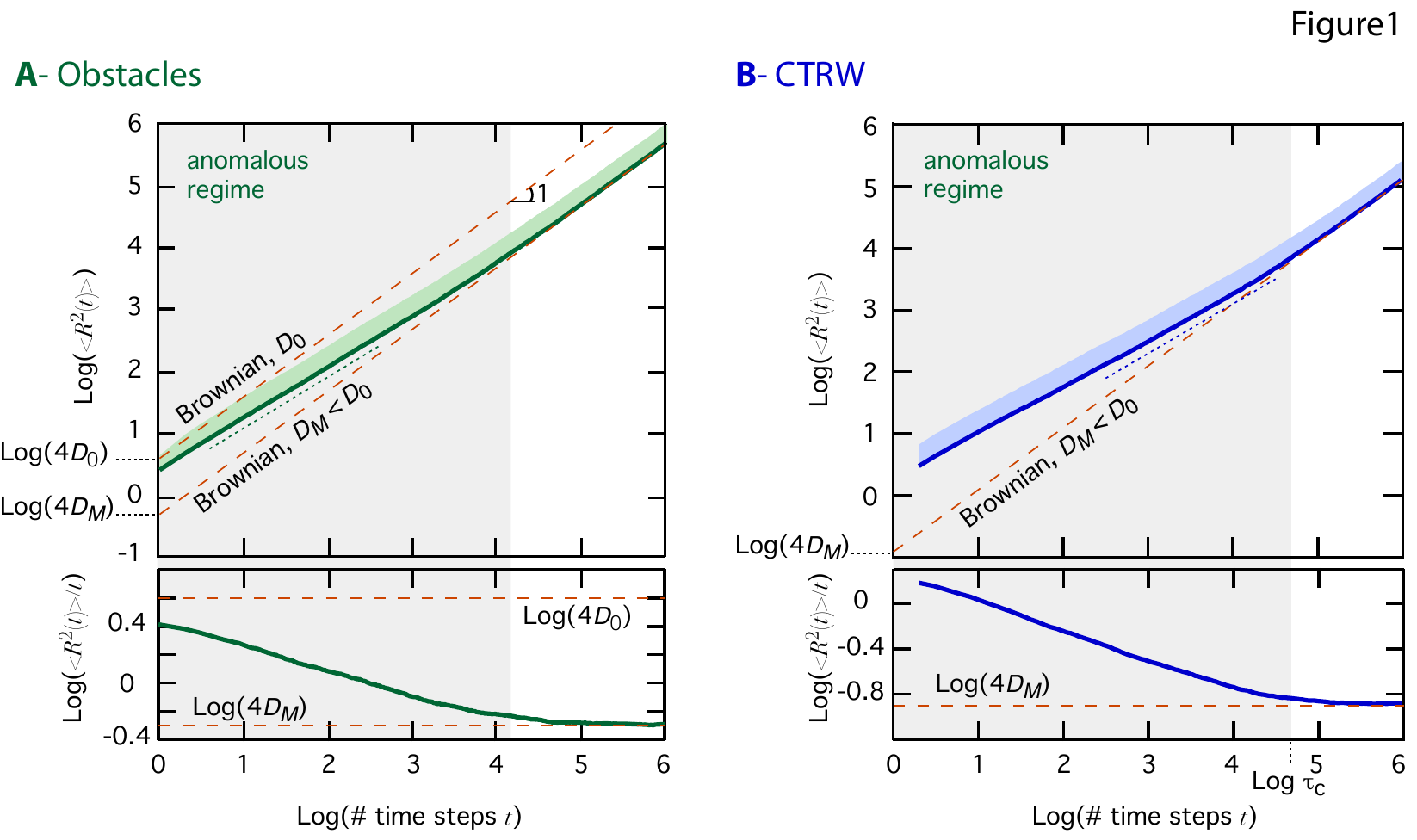}%
 \caption{\label{Fig:Fig1}}
 \end{center}
 \end{figure}

However, the significance of anomalous diffusion for equilibrium properties is questionable since in many experimental data~\cite{Platani2002,Murase2004,Bronstein2009,Jeon2011}, the anomalous regime is only transient: at long times, the mean squared displacement crossovers back to normal (Brownian) diffusion, with $\alpha=1$ but a reduced apparent diffusion coefficient. Such transient behaviors are for instance obtained when the density of obstructing obstacles is below the percolation threshold~\cite{Saxton1994,Hofling2006,Spanner2011} or the residence-time is power-law distributed with a cut-off~\cite{Saxton2007,Jeon2011}. Figure~\ref{Fig:Fig1}A illustrates this transitory behavior with a Monte-Carlo simulation of 2d random walks on a square lattice in the presence of immobile obstacles (obstacle density $\rho=0.35$). For very short simulation times, the distance travelled by the molecules is less than the typical distance between obstacles, so that the movement converges to a Brownian motion without obstacles (with microscopic diffusion coefficient $D_0$). The movement then crossovers to the anomalous subdiffusive regime at longer times, with the mean-squared displacement $\left \langle R^2(t)\right\rangle$ scaling sub-linearly with time (roughly  $\sim t^{0.8}$ in the figure). The anomalous regime however is transitory since, at longer times, the movement crossovers to a second Brownian regime, with a smaller apparent macroscopic diffusion coefficient, $D_M$.  A similar behavior is observed when molecule movements are due to a continuous-time random walk (CTRW), in which the residence time $\tau$ between two successive jumps has a power-law distribution ($\phi(\tau)\propto \tau^{-(1+\alpha)}$ with $0<\alpha<1$)~\citep{Bouchaud1990,Metzler2000}. When the residence time is upper-bounded by a cutoff $\tau_c$ (Figure~\ref{Fig:Fig1}B), the (ensemble-averaged) mean-squared displacement scales anomalously ($\left \langle R^2(t)\right\rangle \propto t^\alpha$) for $t<\tau_c$ then crossovers to a Brownian regime with reduced diffusion coefficient at longer times. In both cases in Figure~\ref{Fig:Fig1}, the transient anomalous behavior transforms to a slowed-down Brownian motion at long times.  This asymptotic slowed-down Brownian regime could be considered a macroscopic (homogenized) representation of the underlying microscopic anomalous diffusion. Following this line of reasoning, it is tempting to assume that the long time (or equilibrium) behavior of a molecule undergoing transient anomalous diffusion can be captured by a slowed-down brownian motion.

Here, we questioned the validity of this assumption, namely, that slowed-down Brownian motion could capture transient anomalous diffusion at long times. We studied the equilibrium properties of the ubiquitous ligand-binding equilibrium eq.\eqref{eq:LR} when diffusion is transiently anomalous either due to obstacles (below the percolation threshold) or to power-law distributed residence times, or when normal space-dependent Brownian diffusion takes place. Using Monte-Carlo simulations and theoretical arguments, we show that this approximation fails even for equilibrium properties, if diffusion conditions are heterogeneous in space.

\section*{Methods}
\subsection*{Brownian motion}
To simulate diffusion, we initially position L and R molecules uniformly at random on a $S=w \times w$ 2D square lattice with reflective boundaries. Each lattice site $(i,j)$ is associated with a diffusion coefficient $D(i,j)$ (all molecules here have identical diffusion coefficients). At each time step $\Delta t$, every molecule is allowed to leave its current location $(i,j)$ with jump probability $\beta(i,j)=4\Delta t / (\Delta x)^2D(i,j)$, where $\Delta x$ is the lattice spacing. The destination site is chosen uniformly  at random from the 4 nearest neighbors $(i\pm1,j\pm1)$ and the molecule jumps to it. For simulation of spatially heterogeneous diffusion, we position the boundary of the slowed-down patch in the middle of neighbor lattice sites. Each lattice side therefore belongs either to the slowed-down patch (we thus set its diffusion constant to $D(i,j)=D_1$) or to the outer region (and we set $D(i,j)=D_0>D_1$).

If the jump of the molecule to its destination site results in the formation of a (L,R) couple on the same lattice site, a binding event may occurs, i.e. the (L,R)  couple is replaced by a single C molecule at the site, with probability $ p_{\rm on}$. Finally, at each time step, every C molecule can unbind, i.e. the C molecule is replaced by a (L,R)  couple at the same site, with probability $ p_{\rm off}$.

\subsection*{Immobile obstacles}
To simulate anomalous diffusion due to obstacles, we position obstacles at random locations (with uniform distribution) at the beginning of the simulation. Obstacles behave a separate type of molecules that are kept unreactive and immobile, while the other molecules (L, R and C) move as indicated above. Obstacles exclude the lattice site they occupy: when the destination site of a moving L, R or C molecule contains an obstacle, the molecule is reflected back to its origin site (the destination site becomes the origin position). Reaction is modeled as for Brownian motion above.

\subsection*{CTRW}
Molecule motion by CTRW is modeled as for Brownian motion above except that upon each jump to its destination site, the molecule is attributed a new residence time $\tau$ sampled from the power-law distribution $\phi(\tau)= \alpha \tau^{-(1+\alpha)}/\left(\Delta t^{-\alpha}-\tau_c^{-\alpha}\right) $, for which $\int_{\Delta t}^{\tau_c}\phi(\tau) d\tau=1$. Hence $\Delta t$, the simulation time step is the smallest residence time possible and $\tau_c$, the cut-off time, sets its maximal value. The next jump of this molecule won't therefore take place before $\tau$ time units are elapsed. Reactions are modeled exactly as for the Brownian case above, with the additional property that molecules can react during residence (i.e. between jumps, whenever they are located at the same location). Moreover every new molecule resulting from a reaction samples a new residence time. Since microscopic details can have crucial effect in CTRW-based reactions~\cite{Henry2006,Yuste2010,Mendez2010}, we have checked in a subset a simulations that the latter does not impact qualitatively our simulation results. 

\subsection*{Simulation Parameters}
In a typical simulation, we start with $r(0)=r_{T}$ R molecules and $l(0)=l_{T}$ L molecules (where $x_{T}$ refers to the total number of molecules X), and no C, and run the simulation until the density of bound receptors C reaches a steady state, $C_{\rm eq}$.
Standard parameter values were used throughout the article, unless otherwise specified: lattice size $w=800$, $r_{T}=100$, $\Delta t=1$, $\Delta x=2$, $ p_{\rm on}=0.1$, $ p_{\rm off}=10^{-3}$ and diffusion coefficient $D_{0}=1$. The ligand dose, $l_T$ was varied so as to obtain dose-response curves. Data were averaged over 20 independent simulations. 

Depending on simulation conditions, equilibrium was typically reached after at most $10^5$ (obstacles) to $5 \times10^5$ (slowed down Brownian diffusion) time steps. The equilibrium value of $C_{eq}$ was therefore computed as the time-average of $C(t)$ for $t \in [4.5,5.0 ]\times 10^5$ (obstacles) or $[9.5,10.0]\times10^5$ (slowed-down Brownian). With CTRW, the time needed to establish equilibrium is much longer than the cutoff $\tau_c$. In all our simulations for $\tau_c\leq10^5$, we observed that equilibrium was reached before $t=9.5\times10^5$ so that we used the values for $t \in [9.5,10.0]\times10^5$ to compute the equilibrium value.

\section*{Results}
\subsection*{Reaction in spatially homogeneous conditions}
We first studied eq.~\eqref{eq:LR} in spatially homogeneous conditions, i.e. in conditions where the diffusion coefficient or local obstacle density is the same everywhere in space.            

\begin{figure}[h!]
  \begin{center}
\includegraphics[scale=1.0]{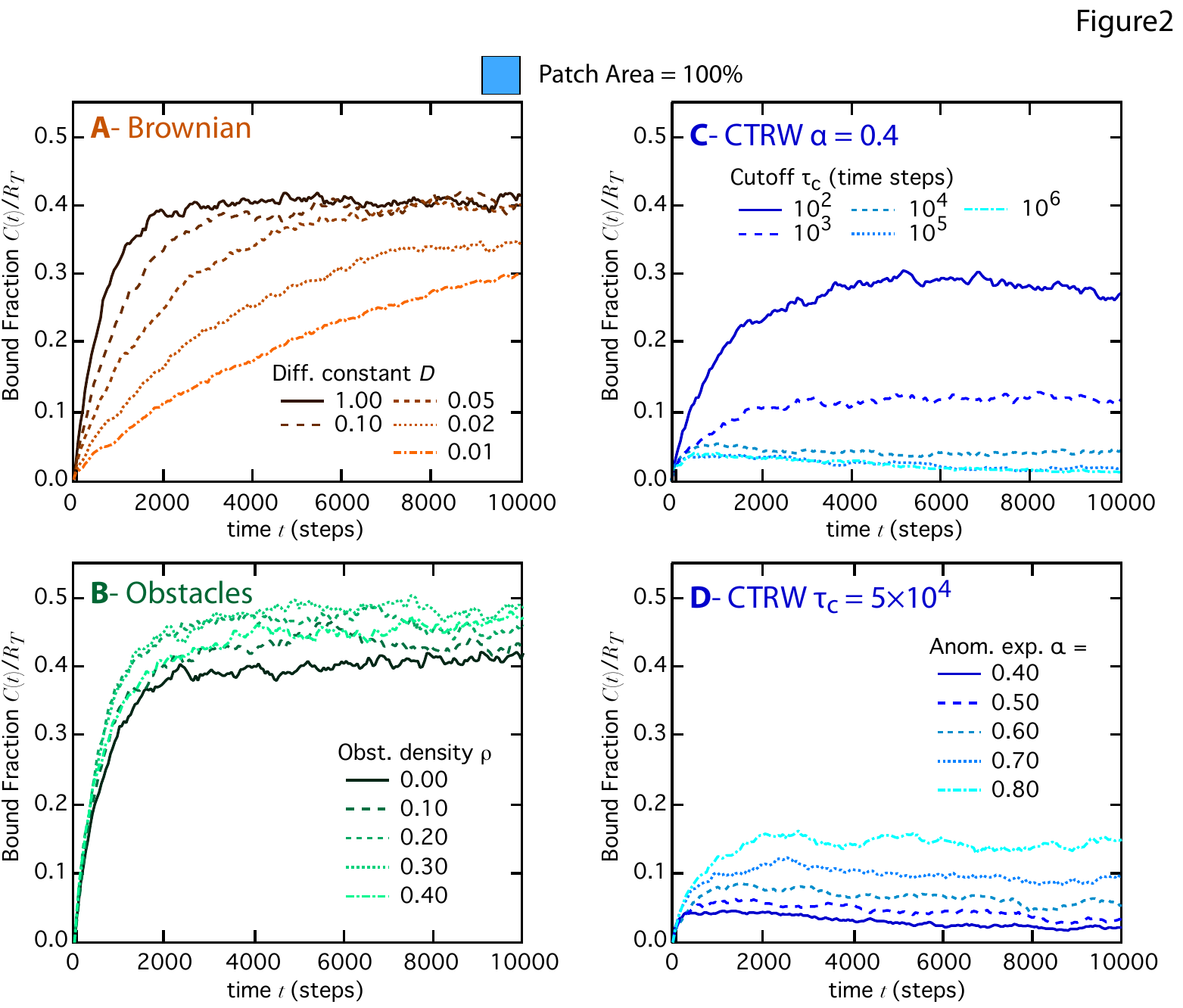}%
 \caption{\label{Fig:Fig2}}
  \end{center}
 \end{figure}

Figure~\ref{Fig:Fig2}A shows typical time-courses of the bound fraction $C(t)/R_T$ for different values of the diffusion coefficient $D$ (no obstacles). While the time needed to reach equilibrium increases with smaller diffusion coefficients, all curves seem to converge at long times to similar levels, thus suggesting that the equilibrium concentration of bound receptor $C_{eq}$ does not depend on the diffusion coefficient. This is of course an expected result from standard thermodynamics: equilibrium configurations should in principle be independent of dynamics, i.e., values of transport coefficients such as diffusion coefficients. The situation is different when the molecule movement exhibits transient anomalous diffusion due to immobile obstacles randomly spread over the whole lattice. At long times (Figure~\ref{Fig:Fig2}B), the reaction as well converges to equilibrium. In this case though, the convergence time to this equilibrium does not seem affected by the density of hindering obstacles, but the concentration of bound receptor at equilibrium seems to vary with obstacle density in a non-trivial fashion. 

\begin{figure}[h!]
  \begin{center}
 \includegraphics[scale=1.0]{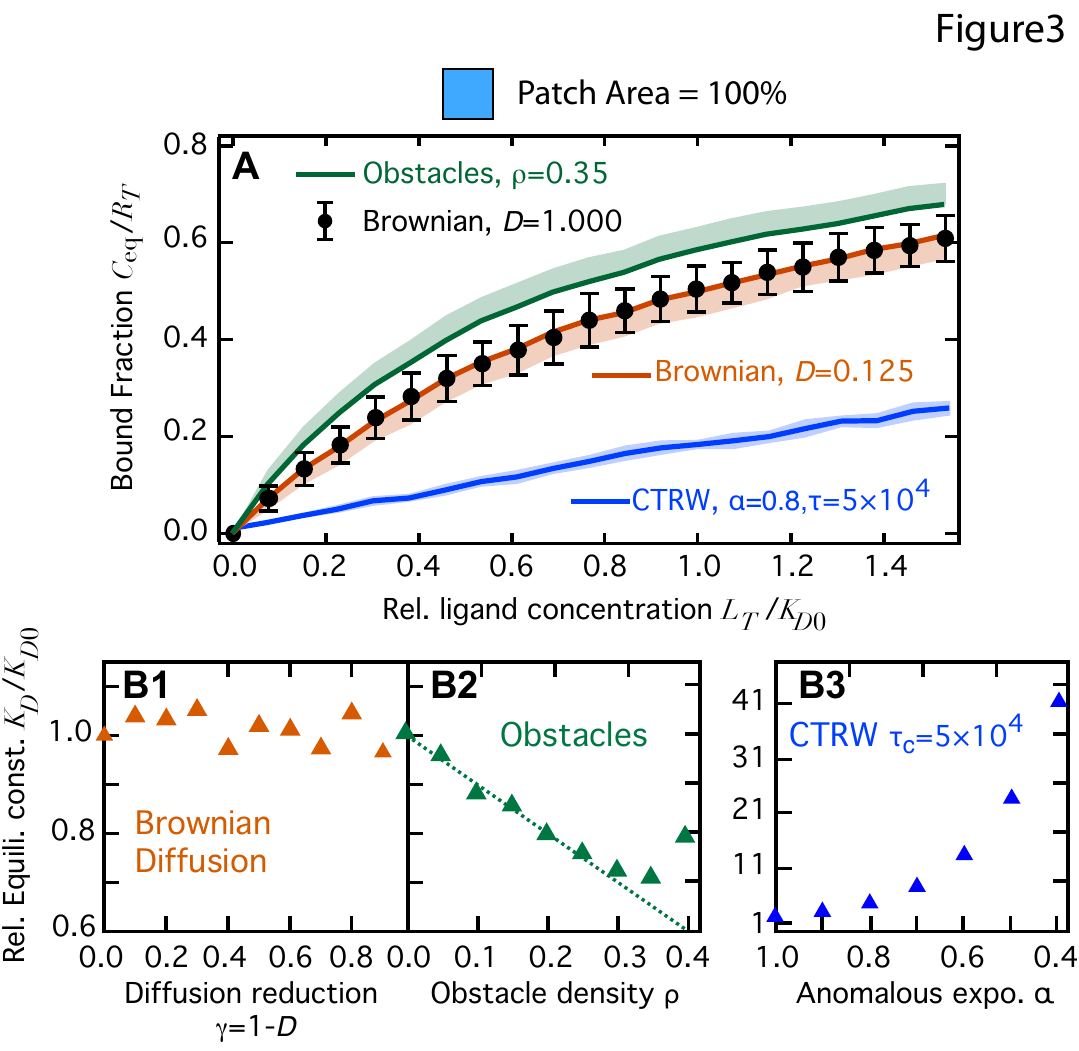}%
 \caption{ \label{Fig:Fig3}}
  \end{center}
 \end{figure}

Figure~\ref{Fig:Fig3}A shows the dose-response curve for an obstacle density $\rho=0.35$ (upper green thick line), for which molecule motion exhibits the transient anomalous diffusive behavior due to obstacles shown in Figure~\ref{Fig:Fig1}A (green thick line). With immobile obstacles, the bound fraction for all doses is found significantly larger compared to the dose-response curve obtained in the absence of obstacles (full black circles). This confirms the observation of Figure~\ref{Fig:Fig2}B that obstacle hindrance alters the bound fraction at equilibrium. Since the molecule movement for $\rho=0.35$ converges at long times to Brownian diffusion with effective macroscopic diffusion coefficient $D_M=0.125$ (Figure~\ref{Fig:Fig1}A) we compared these results with the dose-response curve obtained when the molecules move by a Brownian motion (no obstacles) with diffusion coefficient $D=0.125$ (orange thick line). In agreement with the observation made above, and standard thermodynamics, the corresponding dose-response curve was not significantly different from the curve obtained with $D=1$ (black full circles). This confirms that the macroscopic slowed-down Brownian regime reached at long times during transient anomalous diffusion does not adequately account for equilibrium properties of eq.~\eqref{eq:LR}. 

The dose-response curves for anomalous diffusion due to obstacles maintain a shape that is compatible with the classical dose-response equation ($C_{eq}/R_T=L_T/(K_D+L_T)$). Therefore, we can fit them using this equation and retrieve for all obstacle densities the corresponding apparent equilibrium constant $K_{D}$. Figure~\ref{Fig:Fig3}B  displays $K_{D}$ values for several values of the diffusion coefficient reduction $\gamma=1-D$ in the absence of obstacles (B1) and for several obstacles densities $\rho$ (B2). As expected, even one order of magnitude span for $D$ in the Brownian case does not influence the apparent $K_{D}$ (B1). 

The situation is different for transient anomalous diffusion, though. Far from the percolation threshold ($\rho = 0.41$), $K_{D}$ decays linearly with obstacle density as $K_{D}/K_{D0}=1-\rho$. This is a simple effect of the excluded volume occupied by the obstacles. Indeed, for a constant number of molecules, the available space decreases when obstacle density increases. Consequently, the local density of molecules increases with obstacle density. This gives rise to the measured decrease of the apparent constant $K_D$. In agreement, the decay of $K_D$ far from the percolation threshold disappears if the concentrations are computed on the basis of the accessible space $1-\rho$, instead of the whole space. Therefore, hindered diffusion due to obstacles not only decreases molecule mobility, it also increases the affinity ($\sim 1/K_{D}$) of the reaction. This trend however reverses close to the percolation threshold, where $K_D$ increases. This behavior is due to the competition between two effects: with increasing obstacle densities, the mean first-collision time increases since the macroscopic diffusion coefficient is slower (the first encounter between two distant molecules takes increasingly longer). On the other hand, the re-collision time decreases because re-collisions imply molecules that are initially close by (as a consequence of failed binding attempts or unbinding events) and anomalous diffusion favors re-collisions, see e.g.~\cite{Saxton2002,Guigas2008}. Close to the percolation threshold, the increase of the first-collision time overcompensates by far the decrease of the re-collision time (not shown). As a result, the forward reaction rate $k_{\rm on}$, and the apparent affinity, strongly decrease close to the threshold. 

When anomalous diffusion is due to CTRW, the kinetics of the reaction shows a very different picture (Figure~\ref{Fig:Fig2}C \& D). For anomalous diffusion due to obstacles, obstacle density sets both the (apparent) scaling of the MSD with time in the anomalous regime and the duration of this regime. In the CTRW case though, both quantities (the anomalous exponent $\alpha$ and the crossover time $\tau_c$) are parameters that we  can fix separately. Like for obstacle-induced anomalous diffusion, the convergence time is not much affected by the value of the cutoff time (Figure~\ref{Fig:Fig2}C) nor that of the anomalous exponent (Figure~\ref{Fig:Fig2}D). However, the concentration of bound receptors at equilibrium varies widely with the CTRW parameters.  In general, the equilibrium values with CTRW are much lower than those observed with Brownian motion and obstacle-induced anomalous diffusion. This reduction of equilibrium binding by CTRW progressively attenuates as the cutoff time decays to very low values (Figure~\ref{Fig:Fig2}C) or the anomalous exponent increases (Figure~\ref{Fig:Fig2}D), i.e. when diffusion is increasingly less anomalous and the motion tends to Brownian. Note that the largest cutoff used in this figure ($\tau_c=10^6$) equals the total simulation time, so that, in effect, $\tau_c=10^6$ corresponds to a permanent CTRW regime (no crossover back to the Brownian regime within the simulation time). In this case, equilibrium cannot be reached during the simulation time (CTRW is then a non-equlibrium process).

This strong reduction of equilibrium binding with CTRW is even more obvious in the dose-response curve Figure~\ref{Fig:Fig3}A. Note that the parameters for CTRW in this panel (blue curve) are those illustrated in Figure~\ref{Fig:Fig1}B (blue thick line). For a given ligand dose, the bound fraction at equilibrium with CTRW is much smaller (up to roughly twofolds) than the response curve with Brownian motion, whatever the slow-down (orange thick line). Here as well, since the overall shape of the CTRW dose response curves is compatible with the classical form ($C_{eq}/R_T=L_T/(K_D+L_T)$), they can be fitted to estimate the apparent equilibrium constant $K_{D}$. Figure~\ref{Fig:Fig3}B3 shows that when diffusion becomes increasingly anomalous (the anomalous exponent $\alpha$ decreases from 1.0 downwards), the equilibrium constant increases up to very high values. Therefore, the ligand binding reaction at (long time) equilibrium with CTRW-based or obstacle-based transient anomalous diffusion appears incompatible with slowed-down Brownian motion. However, contrarily to obstacle-based anomalous diffusion, CTRW-based anomalous diffusion strongly impairs the bound fraction at equilibrium and more generally, the affinity of the ligand-binding equilibrium itself.

\subsection*{Space-dependent Brownian diffusion yields accumulation at equilibrium}

The results presented so far consider spatially homogeneous conditions, i.e. $D(i,j)\equiv D, \,\forall i,j$. But in living cells, the conditions are usually spatially heterogeneous : spatial domains (lipid rafts, caveolae) give rise to position-dependent values of $D$
In the following, we addressed this situation by restricting the region of space where diffusion is modified to a central square patch of variable spatial extent. We simulated spatial arrangements where diffusion is Brownian with coefficient $D_{0}$ outside the patch and reduced within the central patch 
by imposing a reduced diffusion coefficient $D_{1}$ inside the patch.

\begin{figure}[h!]
  \begin{center}
 \includegraphics[scale=1.0]{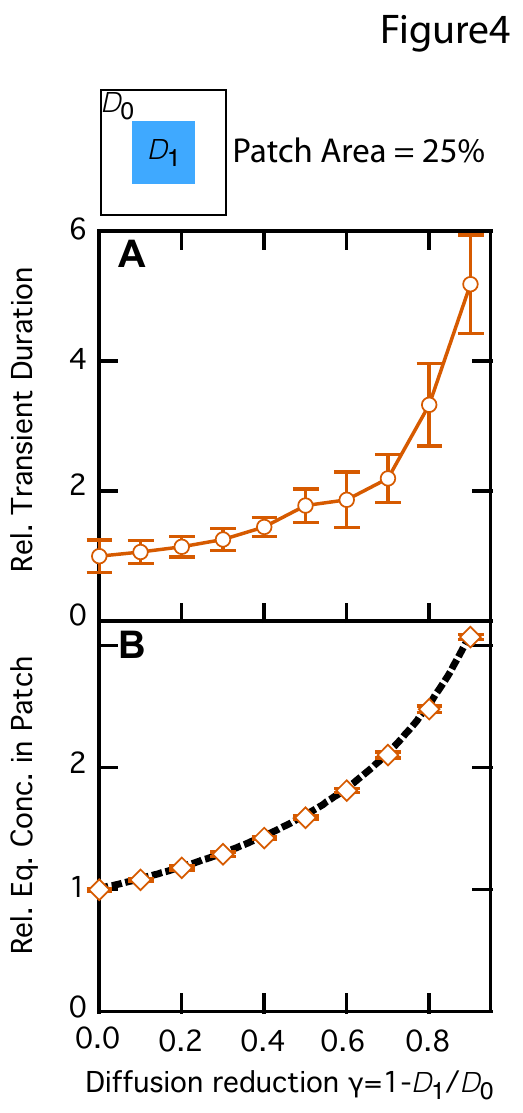}%
 \caption{\label{Fig:Fig4}}
 \end{center}
 \end{figure}

Before addressing the ligand-binding equilibrium eq.~\eqref{eq:LR} in these conditions, we first investigate diffusion, in the absence of reaction. We use a central patch which surface area is 25\% of the whole space and simulate the diffusion of non-reactive molecules until they reach equilibrium. Once equilibrium is reached, we perturb it by the addition of supplementary non-reactive molecules in the center of the patch and measure the characteristic time to reach a new equilibrium and the concentration of molecules inside the patch at this new equilibrium. Because reduced diffusion in the patch slows down the molecules, the characteristic time to converge back to equilibrium increases when diffusion is reduced in the patch (Fig.~\ref{Fig:Fig4}A).
More intriguingly, measuring the concentration of molecules in the patch (relative to the exterior) at equilibrium, we observe increasing accumulation of molecules within the patch when diffusion is slowed-down therein (Fig.~\ref{Fig:Fig4}B). We emphasize here these are equilibrium conditions.

However surprising, this equilibrium effect can be directly predicted in our system.  A first intuitive approach is obtained from the detailed balance condition. A condition for our system to reach (thermodynamic) equilibrium it to respect detailed balance. Consider two states $A$ and $B$ of a Markov process. Note $\eta_A$ the probability to observe state $A$ and $\pi(A \to B)$ the transition probability from $A$ to $B$, the detailed balance condition reads $\eta_A \pi(A \to B) = \eta_B \pi(B \to A)$. Consider now two lattice sites $s_{\rm patch}$ and $s_{\rm out}$ located on either sides of the frontier separating the central patch from the rest of the lattice. The detailed balance in this case reads $\rho(s_{\rm patch})/\rho(s_{\rm out}) =  \pi(s_{\rm out} \to s_{\rm patch})/\pi(s_{\rm patch} \to s_{\rm out})$ where $\rho(x)$ is the concentration of molecules at node $x$. To emulate position-dependent diffusion (see Methods sections), our simulation algorithm states that the jump probability between two lattice sites exclusively depends on the diffusion coefficient at the node of origin. The detailed balance thus becomes $\rho(s_{\rm patch})/\rho(s_{\rm out}) = D_0/D_1 > 1$ (where $D_0$ and $D_1$ are the diffusion coefficients outside and inside the patch, respectively). This predicts accumulation inside the patch at equilibrium. A more formal approach can be applied, based on the master equation. This approach, detailed in the Supporting Material (section A),
predicts that the total number of molecules at equilibrium in the patch $N_{\rm inside}$ relates to total number $N_{\rm total}$, the surface fraction of the patch $\phi$, the total surface $S$ and the diffusion coefficient according to: 
\begin{equation}\label{eq:ninside}
N_{\rm inside}=S\phi N_{\rm total} \frac{\mathcal{H}(D)}{D_1} \text{,\qquad} \mathcal{H}(D)=\left[\frac{\phi}{D_1}+\frac{1-\phi}{D_0}\right]^{-1}
\end{equation}
The theoretical prediction of Eq.~\eqref{eq:ninside} was tested against our Monte-Carlo simulation results in Figure~\ref{Fig:Fig4}B. The agreement between theoretical prediction (thick dashed line) and simulation results (open diamonds) is very good. This confirms that slowed-down Brownian motion in the patch leads to larger concentration inside the patch than outside at equilibrium.

Note that our diffusion algorithm, where the jump probability depends only on the diffusion coefficient at the node of origin, corresponds to solving the Brownian motion with Ito's stochastic calculus. Using Stratonovich's rules instead would preserve accumulation within the patch but with reduced intensity, see~\citet{Schnitzer1993,Soula2012}. To choose which algorithm is the correct one necessitates the knowledge of the microscopic quantities that cause the observed change of diffusion coefficient at macroscopic scales~\cite{Schnitzer1993}~\cite[p.~279-281]{vankampen2007}.

\subsection*{Reaction in spatially heterogeneous conditions}

The accumulation phenomenon described above is likely to modify the reaction in the spatially heterogeneous diffusion case. Using several area fraction for the patch we computed the values of the apparent equilibrium constant $K_{D}$. With anomalous diffusion due to obstacles (Fig.~\ref{Fig:Fig5}A), the behavior reported in Fig.~\ref{Fig:Fig3}B2 is roughly conserved for all patch area fractions $\phi$: $K_{D}$ decreases linearly with the obstacle density $\rho$ far from the percolation threshold then increases back close to it. The amplitude of this decay increases with the patch area (as the total space occupied by obstacles increases). Considering the behavior observed in spatially homogeneous conditions (Fig.~\ref{Fig:Fig3}B2), one expects, far from the percolation threshold, $K_{D}=K_{D0}(1-\phi)+K_{D0}(1-\rho)\phi$, yielding $K_{D}/K_{D0}=1-\rho\phi$. Indeed, we found that the latter is a very good approximation for the simulation results of Fig.~\ref{Fig:Fig5}A. Close to the percolation threshold, the the apparent affinity reverses and starts decreasing because of the strong increase of the first-collision time (see above).

The situation is quite different for space-dependent Brownian diffusion (Fig.~\ref{Fig:Fig5}B). Whatever the patch area, we also observe that $K_{D}$ decreases, but in this case, this is the result of  the accumulation phenomenon reported above (Fig.~\ref{Fig:Fig4}B). Moreover, in this case, $K_{D}$ exhibits a non-monotonous dependency with respect to the area fraction $\phi$, with a marked minimum. Hence, for a given value of diffusion reduction in the patch, $\gamma = 1-D_1/D_0$, our Monte-Carlo simulations show that there exists an optimal value of the patch surface area that yields the highest affinity (Fig.~\ref{Fig:Fig5}B, inset). Using eq~\eqref{eq:ninside} above for both R and L, this optimal value can be estimated analytically. Our theoretical analysis, given in the Supporting Material (section B), indeed shows the existence of an optimal surface area $\phi^\star$ that maximises the apparent reaction affinity. When the size of the space domain $S \gg (l_TD_0)/(D_1K_{D})$ (which is always valid in the simulations shown in the present article, given $S=800^2$), $\phi^\star$ is predicted to depend on $\gamma$ according to:
\begin{equation}\label{eq:phic2}
\phi^\star\approx\frac{1-\gamma}{2-\gamma}
\end{equation}

In the limit of large slowdowns $\gamma \to 1$, eq.\eqref{eq:phic2} gives $\phi^\star \to 0$: the larger the slowdown, the smaller the optimal patch area. This prediction is in general qualitative agreement with the simulation results of Figure~\ref{Fig:Fig5}B that do not depend in a monotonous way on $\gamma$ and $\phi$ but presents extrema along the $\phi$-axis that shift leftward with increasing $\gamma$. In the limit of no-slowdown $\gamma \to 0$, eq.\eqref{eq:phic2} predicts $\phi^\star=0.5$ but then, in this case, the apparent affinity does not depend on $\phi$ anymore (see the Supporting Material, section B), so that no extremum are observed. For a quantitative test of eq.~\eqref{eq:phic2}, we plotted the relationship between $\gamma$ and the value of $\phi$ that exhibited the smallest $K_D$ in our simulations (Figure~\ref{Fig:Fig5}B, inset, open circles). These values are found to nicely align with the prediction of eq.~\eqref{eq:phic2} (inset, full line), thus validating this prediction quantitatively.

Finally figure~\ref{Fig:Fig5}C \& D show the behavior exhibited when anomalous diffusion is due to a CTRW with cutoff time. In agreement with figure~\ref{Fig:Fig3}, we first remark that for almost all the parameters in these figures, reaction affinity is massively impaired by CTRW, up to 4 to 6 orders of magnitude (note that these curves plot the Log of the relative affinity constant). Even for rather short cutoffs (e.g. $\tau_c = 10^3$ time steps), the apparent affinity of the ligand-binding equilibrium is lower with CTRW than with Brownian motion as soon as the patch in which CTRW occurs is wider than one fourth of the total area. For some parameters, the affinity of the reaction can be larger than Brownian motion. This however corresponds to very small cutoff times, that produce anomalous regimes of very limited duration. In those case, the CTRW in the patch tends to Brownian motion with reduced diffusion coefficient so that the system tends to the space-dependent Brownian case illustrated in figure~\ref{Fig:Fig5}B. On the other hand, CTRW is also found to increase the apparent affinity for large cutoff times but only for small to very small patch area fractions. The biological relevance of these restricted cases is therefore not obvious. Therefore, except for very small patch areas, CTRW-based anomalous diffusion massively impairs the affinity of the ligand binding equilibrium.

To conclude, we have shown that heterogeneous slowed-down Brownian systems exhibit a bonus to patchiness : the minimal value of $K_{D}$ is obtained when the patch occupies a subset of the available space. This is in strong contrast with the equilibrium behavior obtained with transient anomalous diffusion above, where the affinity increases or decreases monotonously with the patch area fraction, depending on the microscopic origin of the anomalous behavior (obstacles or CTRW, respectively).

\section*{Discussion}

This study was motivated by the conception that, in contrast to the celebrated fluid mosaic model~\cite{FluidMosaic72}, the cellular plasma membrane is not a simple two-dimensional liquid that would be made spatially homogeneous by the rapid lateral diffusion of lipids and proteins. Recent progress in time-lapse imaging (including at the single-molecule scale) have evidenced the existence of spatial inhomogeneities that form dynamical hierarchical domains at the mesoscale (fence-pickets compartments, raft domains and protein complex domains)~\cite{Kusumi2011}. The impact of this organization in hierarchical domains on the signaling reactions that take place on the membrane is still  poorly understood. The alteration of the diffusive movements of the proteins in these hierarchical domains may be very significant for signaling~\cite{Grecco2011}. However, it is not yet entirely clear what parameters exactly of the diffusive movements are modified in what domains. One could think of modulations of the diffusion coefficient~\citep{Kenworthy:2004p378,Goodwin:2005p1132,Pucadyil:2006p1150,Fujita:2007p1107,Day:2009p2559}, of the confinement distance~\cite{Kamar2012} or of a local change from Brownian to non-Brownian diffusion (anomalous diffusion) due to macromolecular crowding or obstacle hinderance~\cite{Feder1996,Smith1999,Rajani2011,Vrljic2002}. Whether these different scenarios have different effects on biochemical reactions on the membrane is not known.    

In this article, we focused on the comparison between three of these scenarios: Brownian diffusion with reduced diffusion coefficient (slowed-down Brownian) and transient anomalous diffusion due to immobile obstacles or power-distributed residence times (CTRW). Indeed, since transient anomalous diffusion converges at long time to a slowed-down Brownian movement, one may be led to consider slowed-down Brownian motion as equivalent to transient anomalous diffusion at equilibrium. In contrast, we have shown here that this assumption fails when diffusion conditions are spatially heterogeneous since the equilibrium behavior of the three scenarios we contemplated are markedly distinct:  when it is due to obstacles, transient anomalous diffusion increases the apparent binding affinity, with a maximal effect when the obstacles are spread all over the available space. Whereas when it is based on a CTRW, transient anomalous diffusion strongly decreases the apparent binding affinity. Slowed-down Brownian motion has a very different effect, since it increases the apparent affinity in a non-monotonous way: maximal affinity is reached when the region of reduced diffusion coefficient is restricted to a subdomain of the membrane surface. Therefore, slowed-down Brownian motion in the ligand binding reaction does not capture the effect of transient anomalous diffusion even at the long times necessary to reach equilibrium. 

A main result from our study is that CTRW and hinderance by immobile obstacles lead to very different behaviors at equilibrium although they yield comparable anomalous scaling of the mean-squared displacement. This result is in line with the realization that the two processes are fundamentally distinct. For instance, CTRW present a weak ergodicity breaking (scaling with time of the time-averaged MSD differs from that of ensemble-averaged MSD)~\cite{Burov2011,Jeon2011} that is not observed with obstacle-based anomalous diffusion. The scaling with time or initial distance of several observables derived from first-passage time statistics differ notably between the two processes~\cite{Condamin2008}. Whether the distinct equilibrium behaviors disclosed in our study are related to these differences is currently unknown but will be investigated in future works. Paradoxically however, these two processes need not be mutually exclusive but may coexist. For instance it has recently been suggested that the motion of ion channels on the cytoplasmic membrane would be consistent with a CTRW that is restricted to take place on a fractal~\cite{Weigel2011}. Since hindering by randomly located immobile obstacles restricts the walker movement to such a fractal geometry (at least close to the percolation threshold)~\cite{Saxton1994}, such a phenomenon could in principle be studied in our simulations. However, adding the slowdown of the reaction due to obstacles with that due to CTRW may be problematic in terms of simulation times and demand alternative simulation or modeling frameworks.

Fractional Brownian motion (and the associated fractional Langevin equation) has recently been evidenced as a third possible source of anomalous diffusion, in addition to obstacle hindering and CTRW.  Fractional Brownian motion (fBM) is a generalization of classical Brownian motion, where the random increments between two successive locations are not independent (like in Brownian motion) but present long-range temporal correlations~\cite{barkai-phystoday-2012}. Like CTRW and hindering by immobile obstacles, fBM gives rise to anomalous diffusion but no weak ergodicity breaking. Most notably, fBM could play an important role in the diffusion of lipids in membranes and be a major source of anomalous diffusion therein~\cite{kneller2011,Jeon2012,Javanainen2013}. Note however that fBM has also been proposed to describe the long-time regime in the transport of lipid granules in \textit{S. pombe}~\cite{Jeon2011}. The impact of fBM on (bio)chemical reactions and in membrane signaling in particular has not yet been thoroughly studied (see however~\cite{Hellmann2012}) but may become an important topic if the implication of fBM in lipid movements in membranes is confirmed.  

Our results for slowed-down Brownian motion suggest that in the case where the membrane is partitioned into two regions only (the patch, into which diffusion is slowed-down and the rest of the lattice), a surface area for the patch exists that optimizes the apparent reaction affinity. This is however a very simplified configuration, since their may exists several disconnected (slowed-down) patches coexisting in the membrane. It is unknown whether in this case, a (total) optimal surface area would still exist or what type of spatial configurations of the patches would be optimal (if ever). In terms of combinatorics, the numerical study of this problem by Monte Carlo simulations would be very challenging since the number of configurations for a given total patch surface area is very large, but could reveal very interesting properties regarding space-dependent Brownian motto.

The functional implications of our finding may be significant for our understanding of the organization of cell membranes, and more generally, cell spaces. For instance, it is very attractive to remark that, in living cell membranes, slowed-down regions (e.g. rafts) show a very patchy distribution, whereas bulky obstacles seem less systematically clustered in limited regions. Controlling the spatial extension of the areas with reduced lateral diffusion may thus be a way by which cells control the apparent affinity of the ubiquitous ligand-reaction binding events.

\section*{Acknowledgments} This research was supported by INRIA grant ``AE ColAge'' and a fellowship from Rh\^{o}ne-Alpes Region to B.C.

\bibliography{Soulaetal_2013}

\clearpage
\section*{Figure Legends}

\subsubsection*{Figure~\ref{Fig:Fig1}. Transient anomalous diffusion as a transitory behavior to a macroscopic slowed-down Brownian regime.} Time-evolutions (Log-Log scales) of the mean-squared displacement, $\left \langle R^2(t)\right\rangle$ (top) and corresponding evolution of the ratio $\left \langle R^2(t)\right\rangle/t$ (bottom) during transient subdiffusive anomalous diffusion due to obstacle hinderance (\textit{A}) or power-law distributed residence times (CTRW) (\textit{B}).  Thick lines show the (ensemble) average while the light swaths indicate +1 s.d. In the top panels, Brownian motion manifests as a straight line with unit slope and a $y$-intercept set by the diffusion coefficient (thin dashed orange lines). The anomalous regime is observed as a transient behavior, with slope $\alpha \approx 0.80$ in (\textit{A}) or $\alpha = 0.80$ in (\textit{B}) (dotted lines), crossing over to an effective macroscopic Brownian regime with diffusion coefficient $D_M$. For panel (\textit{A}), one gets $D_M\approx 0.125$ whereas $D_M\approx 0.032$ in (\textit{B}). In the bottom panels, Brownian motion manifests as a horizontal straight line with $y$-intercept set by the diffusion coefficient and the anomalous regimes as straight lines with slope $1-\alpha$. Parameters: $\Delta t=1$, $\Delta x=2$ and $D_{0}=1$, domain size $w=10^6$. Data are averages of $10^4$ independent trajectories (\textit{A}) obstacle density $\rho=0.35$, (\textit{B}) CTRW exponent $\alpha=0.8$, cutoff time $\tau_c=5\times10^4$.
anomalous diffusion

\subsubsection*{Figure~\ref{Fig:Fig2}. The transient dynamics of reaction reaction eq.~\eqref{eq:LR} in homogeneous conditions, $D(i,j)\equiv D, \,\forall i,j$.} The time-evolution of the bound fraction $C(t)/R_T$ is shown either ({\it A}) for values of $D$ decreasing from 1.0 to 0.01, from top to bottom, respectively (no obstacles), ({\it B}) for obstacle densities increasing from 0.0 to 0.40 (with microscopic diffusion coefficient $D_0=1$),  ({\it C}) for CTRW motion with $\alpha=0.4$ and cutoff time $\tau_c=10^2,10^3,10^4,10^5$ or $10^6$ (from top to bottom); or ({\it D}) for CTRW motion with cutoff time $\tau_c=5\times 10^4$ and $\alpha=0.4,0.5,0.6,0.7$ or $0.8$ (from bottom to top) Note that since the total simulation time is $10^6$ time steps, $\tau_c=10^6$ corresponds to a CTRW with permanent anomalous regime (no crossover back to Brownian during the simulation). Total ligand number $l_T=4500$ and all other parameters were set according to the standard set (see Methods). 

\subsubsection*{Figure~\ref{Fig:Fig3}. Equilibrium study of eq.~\eqref{eq:LR} in homogeneous conditions, $D(i,j)\equiv D, \,\forall i,j$.} ({\it A}) The bound fraction at equilibrium, $C_{\rm eq}/R_T$ as a function of the relative ligand dose $L_T/K_{D0}$, where $K_{D0}$ is the value of $K_{D}$ in the absence of obstacles and with reference diffusion coefficient $D_{0}=1$. ($\rho$, $D$)=(0.0, $D_{0}$) (black, bars show $\pm$ 1 s.d.), (0.0, $0.125$) (orange, light swath shows -1 s.d.) or (0.35,1) (green, light swath shows +1 s.d.). The bound fraction at equilibrium for CTRW (with $\alpha=0.8$ and $\tau_c=5\times10^4$) is shown in blue  (light blue swaths show $\pm$1 s.d.). From Student's $t$-tests, the data points in the two Brownian cases are not significantly different (at identical ligand dose), whereas the data points in the presence of obstacles or with CTRW are each significantly different at all doses (except 0) from the Brownian cases (significance level $p<0.01$). ({\it B}) Relative apparent equilibrium constant $K_{D}/K_{D0}$ (B1) without obstacles but increasing reduction of the diffusion, $\gamma=1-D$ or (B2) with $D=1$ but increasing obstacle density $\rho$ or (B3) for CTRW-based motion with decreasing values of the anomalous exponent $\alpha$ (cutoff time $\tau_c=5\times10^4$). The dashed line locates the diagonal $y=1-x$.  Other parameters were set according to the standard set (see Methods). 

\subsubsection*{Figure~\ref{Fig:Fig4}. Brownian diffusion in heterogeneous (space-dependent) conditions.} Nonreactive molecules diffuse with coefficient $D_0$ (no obstacles) outside of a central patch in which diffusion is slowed-down (the diffusion coefficient inside the patch is $D_{1}=D_0(1-\gamma)$). ({\it A}) shows the characteristic time to reach equilibrium while ({\it B}) shows the molecule density at equilibrium in the patch.  Data are normalized by the values obtained in the absence of slow-down $\gamma=0$. The full line in (A) is a guide to the eyes while the black dashed line in (B) shows the theoretical prediction eq.~\eqref{eq:ninside}. Bars indicate $\pm$ 1 s.d. Other parameters were set according to the standard set (see Methods).

\begin{figure}[h!]
 \begin{center}
 \includegraphics[scale=1.0]{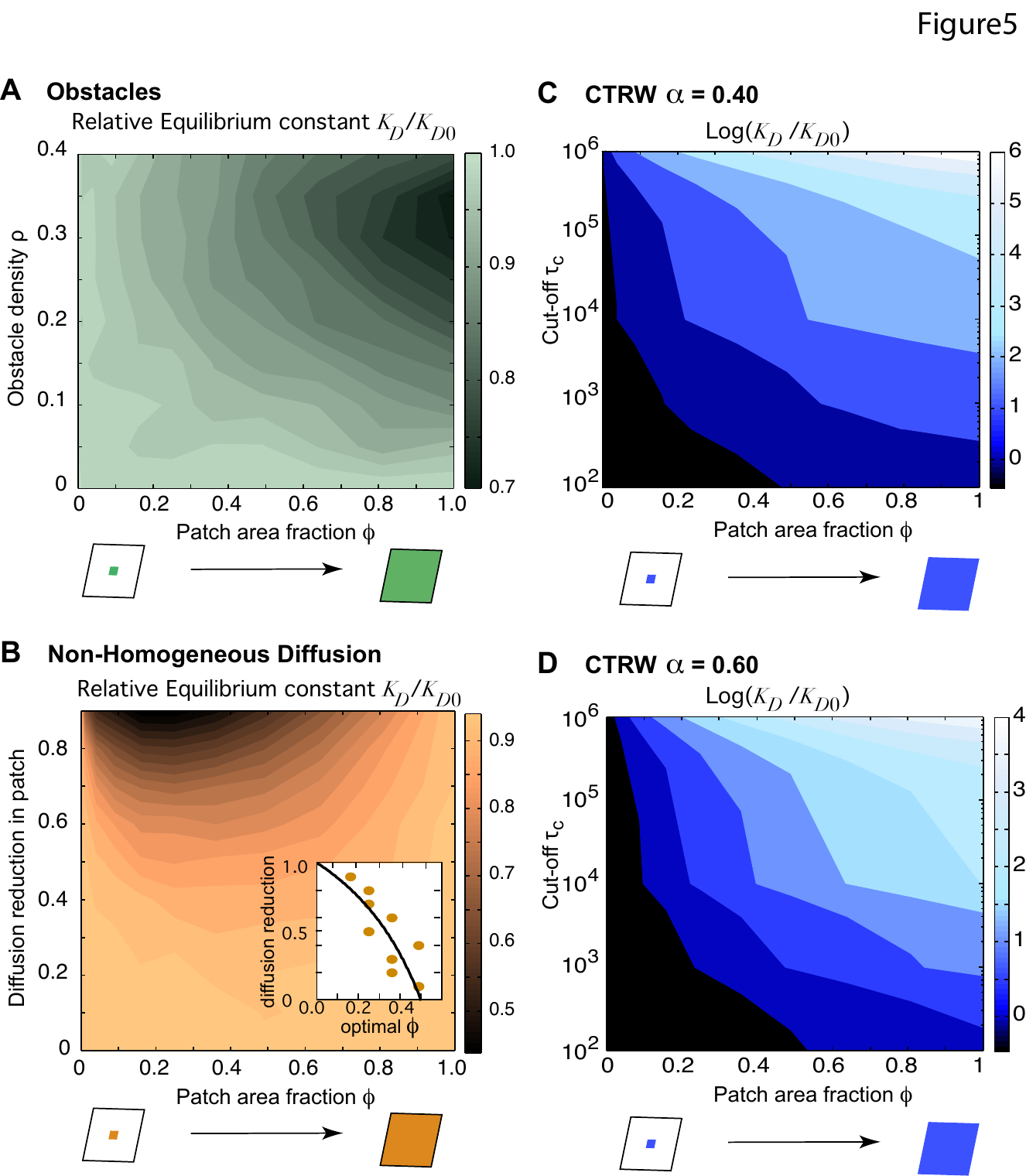}%
 \caption{ \label{Fig:Fig5}}
 \end{center}
 \end{figure}

\subsubsection*{Figure~\ref{Fig:Fig5}. Equilibrium properties of reaction eq.~\eqref{eq:LR} in space-dependent conditions.} Molecules move with Brownian motion outside the central patch. Inside the central patch, molecule motion is due to ({\it A}) transient anomalous diffusion due to obstacles, ({\it B}) slowed-down Brownian motion or CTRW with ({\it C}) $\alpha=0.40$ or ({\it D})  $0.60$. The panels show the apparent equilibrium constant $K_{D}/K_{D0}$ as a function of the area fraction occupied by the patch $\phi$ and the obstacle density $\rho$ ({\it A}), the amount of diffusion reduction in the patch, $\gamma=1-D_1/D_0$ ({\it B}) or the cutoff time $\tau_c$ (({\it C-D}) (Note that (C-D) show  $\log_{10} \left(K_{D}/K_{D0}\right)$). 
 The inset in ({\it B}) locates the value of $\phi$ yielding maximal affinity (full circles) and the corresponding theoretical prediction eq.~\eqref{eq:phic2} (full line). Other parameters were set according to the standard set (see Methods).

\clearpage

 \clearpage
 
\clearpage

\clearpage

\clearpage

 \newpage

\section*{Supplementary Information}
\begin{center}
\texttt{\Large{Soula et al.  \\Anomalous versus slowed-down Brownian diffusion in the ligand-binding equilibrium}}\\\vspace{0.5cm}
\texttt{\Large{Supporting Material}}\\
\end{center}
\renewcommand{\theequation}{SI.\arabic{equation}}
\setcounter{equation}{0}
\subsection*{A. Space-dependent Brownian diffusion yields accumulation at equilibrium}

Let us consider the 1d case for simplicity, and a constant-by-part dependence of the diffusion coefficient $D(x)=D_{1},\;\forall x\in[a,b]$ and $D(x)=D_{0}$ outside the patch $[a,b]$. Let us then consider a single molecule and let $\pi(x,t)$ its probability to be located at position $x$ at time $t$:
\begin{eqnarray}
\pi(x,t+\Delta t)=&&q(x)\pi(x,t)+\pi(x-\Delta x,t)\left(1-q(x-\Delta x)\right)/2\nonumber\\&&+  \pi(x+\Delta x,t)\left(1-q(x+\Delta x)\right)/2
\end{eqnarray}
where $q(x)$ is the probability not to jump at each time step and is defined, using the jump probability $\beta(x)=2\Delta t / (\Delta x)^2D(x)$ (see Methods), as $q(x)=1-\beta(x)$. Noting $g(x,t)=(1-q(x))\pi(x,t)/2$ and developing $g(x\pm\Delta x,t)$ in series of $x$, one obtains at order 2
\begin{eqnarray}
\pi(x,t+\Delta t)&&=q(x)\pi(x,t)+2g(x)+{(\Delta x)}^{2}\partial_{xx}g(x)\nonumber\\
&&=\pi(x,t)+(\Delta x)^{2}\partial_{xx}g(x,t)
\end{eqnarray}
Dividing by $\Delta t$ and taking the limit $\Delta t \to 0$, one gets
\begin{equation}\label{eq:SI_FP}
\partial_{t}\pi(x,t)=\partial_{xx}\left(D(x)\pi(x,t)\right)
\end{equation}
where we used the expression of $\beta(x)$ above to define $D(x)$. Noting $u(x,\infty)$ the density of molecules at $x$ at equilibrium, one expects from eq.~\eqref{eq:SI_FP}
 \begin{equation}
D(x)u(x,\infty)=\mathcal{H}(D)
\end{equation}
where $\mathcal{H}(D)$ is the spatial harmonic mean of the (space-dependent) diffusion function
\begin{equation}
\mathcal{H}(D)=\left[\int D^{-1} (x) dx\right]^{-1}
\end{equation}
Now, using the constant-by-part function for $D(x)$ expressed above, this yields $u(x,\infty)=\mathcal{H}(D)/D_{1}\;\forall x\in[a,b]$ and $u(x,\infty)=\mathcal{H}(D)/D_{0}$ outside. The equilibrium concentration inside the $[a,b]$ patch thus equals that found outside the patch multiplied by $D_{0}/D_{1}$. Hence the larger the slowdown of the Brownian motion inside the patch, the larger the accumulation inside it at equilibrium, explaining the simulation results of Fig.4B. In the present 2d case, the total number of molecules in the patch $N_{\rm inside}$ relates to total number $N_{\rm total}$, the surface fraction of the patch $\phi$, the total surface $S$ and the diffusion coefficient according to: 
\begin{equation}\label{eq:SI_ninside}
N_{\rm inside}=S\phi N_{\rm total} \frac{\mathcal{H}(D)}{D_1} \text{~,~} \mathcal{H}(D)=\left[\frac{\phi}{D_1}+\frac{1-\phi}{D_0}\right]^{-1}
\end{equation}

\subsection*{B. Optimum area for spatially restricted slowed-down Brownian motion}
Let us consider a  space domain of total area $S=w\times w$, in which molecules move by Brownian motion with diffusion coefficient $D(x)=D_1$ inside the central patch (of surface $\phi S$) and $D(x)=D_0$ in the outer region around this central patch (surface  $(1-\phi) S$). We denote numbers of molecules by lower-case letters to distinguish them from concentrations (denoted by capital letters): $x$ thus expresses the number of X molecules in the domain. Moreover, just like for the diffusion coefficient above, we use indices for each variable to indicate location, i.e. $x_1$ refers to the number of X molecules within the central patch while $x_0$ refers to its value outside the patch. Finally, in the following, all results will relate to equilibrium values, so that we drop the ``eq'' notation used above for readability.


Our major assumption in the following theoretical analysis is to consider that the reaction proceeds separately in each zone (inside or outside of the patch), independently of each other. Our goal then becomes to determine the value of $\phi$ that maximizes $c_0+c_1=c_T$, the total number of complexes. According to our space separation assumption, one has in each zone $i=\{0,1\}$:
\begin{equation}
C_i=\frac{R_{i,T} L_{i,T}}{K_{Di} + L_{i,T}}
\end{equation} 
where $R_{i,T}=R_i+C_i$ and $L_{i,T}=L_i+C_i$. In terms of molecule numbers, this translates into 
\begin{equation}
c_0=\frac{r_{0,T} l_{0,T}}{K_{D0}(1-\phi) S + l_{0,T}}\; \mathrm{and\;}
c_1=\frac{r_{1,T} l_{1,T}}{K_{D1}\phi S + l_{1,T}}
\end{equation}

Now, according to eq.~\ref{eq:SI_ninside}, the relative amount of reactants in each zone is given by 
\begin{equation}\label{eq:SI_nhd}
\rho(x)=\frac{\mathcal{H}(D)}{D(x)}
\end{equation}
with $\mathcal{H}$ the (2D) spatial harmonic mean of the diffusion
constant $D$
\begin{equation} \label{eq:SI_harm}
\mathcal{H}(D)=\left[\iint_S D^{-1}(x)dx\right]^{-1}= \left[S\frac{\phi}{D_1}+S\frac{(1-\phi)}{D_0}\right]^{-1}
\end{equation}
The amount of reactant outside the central patch thus reads
\begin{equation*}
r_{0,T}=r_T
\int_{(1-\phi)S}\rho(u)du=\iint_{(1-\phi)S}\frac{\mathcal{H}(D)}{D(u)}du
\end{equation*}
so that
\begin{equation}
r_{0,T}=\frac{r_T \left(1-\phi\right)S\mathcal{H}(D)}{D_0}
\end{equation}
Likewise, inside the patch:
\begin{equation*}
r_{1,T}=r_T
\int_{\phi S}\rho(u)du=\iint_{\phi S}\frac{\mathcal{H}(D)}{D(u)}du
\end{equation*}
yielding 
\begin{equation}
r_{1,T}=\frac{r_T \phi S\mathcal{H}(D)}{D_1}
\end{equation}
Note that 
$$\left(1-\phi\right) S \mathcal{H}(D)/D_0+\phi S \mathcal{H}(D)/D_1
= 1$$ and the above results stands for $l_{i,T}$ ($i=\{0,1\}$) as well. 

Therefore, noting
\begin{equation}\label{eq:SI_alpha}
\alpha=\left(1-\phi\right)S\mathcal{H}(D)/D_0
\end{equation}
and
\begin{equation}\label{eq:SI_oneminusalpha}
1-\alpha=\phi S\mathcal{H}(D)/D_1
\end{equation}
we obtain
\begin{equation}\label{eq:SI_cT}
c_T=c_0+c_1=r_Tl_T\left(\frac{\alpha^2}{K_{D0}(1-\phi)S+\alpha l_T}+\frac{(1-\alpha)^2}{K_{D1}\phi S+(1-\alpha) l_T}\right)
\end{equation}
In particular, in homogeneous conditions ($D_0=D_1$ and $K_{D0}=K_{D1}$), one has $\mathcal{H}(D)=D_0/S$ and $\alpha=1-\phi$ so that eq.\eqref{eq:SI_cT} reduces to $c_T=r_Tl_T/(K_{D0}S+l_T)\; \forall \phi$, i.e. precisely the classical dose-response curve for homogeneous conditions. Note that except for homogeneous conditions, eq.\eqref{eq:SI_cT} does not generally display the classical parabolic shape, typical of the homogenous conditions ($y=cx/(d+x)$). 

Now, the assumption of space separation between the two zones means that the movement is homogeneous (position-independent) Brownian motion for each zone. In this case we have found on Figure~3B1 (main text) that $K_{D1} \approx K_{D0}$ for all values of $D_1$ tested ($D_0=1$). We thus set $K_{D0}=K_{D1} \equiv K_D$ in the following. To find the extremum of eq.\eqref{eq:SI_cT}, we search for the solutions of $dc_T/d\phi=0$ and get:
\begin{equation}\label{eq:SI_phic1}
\phi^\star = \frac{S+ad-\sqrt{d(S+ad)(a+Sd)}}{S (1-d^2)}
\end{equation}
where we noted $d \equiv D_0/D_1$ and $a \equiv l_T/K_{D}$. We remark that in this expression, the value of the optimum area $\phi^*$ depends on the dose, i.e. the total concentration of ligand $L_T=L+R$.   This is related to the fact that eq.\eqref{eq:SI_cT} generally has not a typical parabolic shape. However, eq.\eqref{eq:SI_phic1} greatly simplifies when $S \gg ad$ (which is always valid in the simulations shown in the present article, given $S=800^2$), to a very simple expression

\begin{equation}\label{eq:SI_phic2}
\phi^\star\approx\frac{1-\gamma}{2-\gamma},\quad S\gg ad
\end{equation}

with $\gamma=1-D_1/D_0$. It is remarkable that, in this limit, $\phi^\star$ does not depend on the dose $a$ anymore, which in fact relates to the fact that the expression for $c_T$ (eq.\eqref{eq:SI_cT}) in this case adopts a classical parabolic shape.

Taken together, this simple theoretical analysis predicts the existence of an optimal surface area $\phi^\star$ for the affinity, that depends on the value of $D_1$ relative to $D_0$. In the limit of large slowdowns $\gamma \to 1$, eq.\eqref{eq:SI_phic2} gives $\phi^\star \to 0$: the larger the slowdown, the smaller the optimal patch area. This prediction is in general qualitative agreement with the simulation results of Figure5B that do not depend in a monotonous way on $\gamma$ and $\phi$ but presents extrema along the $\phi$-axis that shift leftward with increasing $\gamma$. In the limit of no-slowdown $\gamma \to 0$, eq.\eqref{eq:SI_phic2} predicts $\phi^\star=0.5$ but then, in this case, the value of $c_T$ does not depend on $\phi$ anymore (see above for $D_0=D_1$), so that no extremum are observed.

\end{document}